\documentclass[final,5p,12pt,twocolumn]{elsarticle}

\usepackage{lineno}
\usepackage{amsmath,amssymb,mathtools,bm,physics,microtype}
\usepackage{siunitx}
\usepackage{float}         
\usepackage{graphicx}
\usepackage{subcaption}
\usepackage{booktabs}
\usepackage{tabularx}
\usepackage[version=4]{mhchem}
\usepackage{xcolor}
\usepackage[normalem]{ulem}

\usepackage{hyperref}
\hypersetup{
  colorlinks=true,
  citecolor=blue,
  urlcolor=blue,
  linkcolor=blue
}

\usepackage{lineno}

\journal{Nuclear Instruments and Methods in Physics Research A}

\begin{document}

\begin{frontmatter}

\title{Adiabatic Fast Passage Spin Manipulation Measurements in Solid Polarized Targets}

\author{M.F. Hossain\corref{cor1}}
\ead{dgy5cd@virginia.edu}
\author{K. Nakano}
\author{N.G. Vismith}
\author{D. Keller}
\address{University of Virginia, Department of Physics, Charlottesville, VA, USA}
\cortext[cor1]{Corresponding author}

\begin{abstract}
Adiabatic fast passage (AFP) is a rapid method for reversing nuclear polarization and manipulating spin populations in polarized solid targets, avoiding the long repolarization times associated with dynamic nuclear polarization (DNP). We report AFP measurements in a 5~T, 1~K polarized-target system for irradiated $^{15}$NH$_3$, irradiated $^{14}$ND$_3$, and butanol-based materials prepared either with TEMPO doping or by irradiation. We also present a joint manipulated-lineshape analysis for spin-1 targets and demonstrate that vector and tensor polarizations can be extracted from AFP-manipulated deuteron NMR spectra even when the populations are not described by a single Boltzmann spin temperature. Finally, we report a reproducible polarization- and direction-dependent AFP response in a large irradiated $^{15}$NH$_3$ sample. These ammonia results are presented as empirical observations under the specific sample--coil conditions of the experiment, with possible circuit-mediated mechanisms such as radiation damping or superradiant behavior discussed but not assigned as a definitive cause. 
\end{abstract}


\end{frontmatter}

\section{Introduction}
Polarized solid targets are central tools for spin-dependent measurements in nuclear and high-energy physics. A recurring operational limitation is polarization reversal: reversal by DNP repolarization is reliable but slow---often requiring hours for some materials---during which beam conditions and detector acceptances may drift. AFP provides a rapid, coherent inversion mechanism by sweeping the RF frequency (or the holding field) through the NMR resonance so that the macroscopic magnetization follows an effective field in the rotating frame \cite{Bloch1946Induction,Abragam1961}. The method is conceptually related to the ``single-shot passage'' and spin-temperature descriptions developed for solids \cite{Goldman1968,Goldman1970,Provotorov1962}.

Historically, AFP performance has shown strong material dependence. Early studies in proton targets with relatively high paramagnetic doping found large losses that limited practical use \cite{Parfenov1984}. In contrast, Hautle \textit{et al.} demonstrated high AFP efficiencies for deuterons in deuterated alcohols (spin-1) \cite{Hautle1992,Hautle1995}. These results highlighted that the relevant optimization for frozen-spin DNP targets is not only the classical adiabatic criterion but also the interplay of line width, relaxation in the rotating frame, and the coupling between nuclear spin reservoirs.

The present work extends AFP measurements and analysis in a modern 5~T/1~K polarized-target system \cite{CrabbMeyer1997}. We first report AFP efficiency measurements in irradiated \ce{^{14}ND3} and \ce{^{15}NH3}, and in butanol systems prepared either with TEMPO doping or by irradiation. We then develop a joint manipulated-lineshape polarimetry technique, demonstrated with irradiated deuterated butanol AFP spectra, that extracts the vector polarization \(P\) and tensor polarization \(Q\) in a spin-1 system directly from manipulated NMR spectra. Throughout the paper, a ``half-flip'' state denotes an interrupted or incomplete AFP manipulation in which the spin-1 sublevel populations are not described by a single global Boltzmann spin temperature; it is an operational description of the manipulated spectrum, not a new equilibrium state. Finally, we report a 7~g irradiated \ce{NH3} study in which the measured reversal efficiency depends reproducibly on initial NMR area and sweep direction under the specific large-sample circuit conditions used here (Table~\ref{tab:nh3_init}).

\section{Experimental apparatus and target materials}
\label{sec:apparatus}

\subsection{5~T/1~K polarized target system}
All measurements were performed in the University of Virginia polarized target laboratory using the same 5~T/1~K DNP target architecture described in Refs.~\cite{CrabbMeyer1997,Keller2020,Pierce2014,Keller2013}. The system consists of a superconducting solenoid providing a uniform holding field, a \(^4\)He evaporation refrigerator capable of providing \(\sim\)1~K operation under beam-equivalent heat loads, and a microwave system for DNP pumping.

The NMR polarimetry employs a series-tuned LCR circuit in the standard Q-meter system \cite{Court1993}, with the NMR absorption obtained from the dispersive response after background subtraction and calibration \cite{Keller2013,Dulya1997}. At 5~T, the NMR frequency is \(\approx\)213~MHz for protons and \(\approx\)32.7~MHz for deuterons.

\subsection{Target cups and sample masses}
For testing purposes our target insert supports two interchangeable cups:
\begin{itemize}
\item A \textbf{small cup} containing approximately \(1~\mathrm{g}\) of granulated material, and
\item A \textbf{large cup} containing approximately \( 7~\mathrm{g}\) of granulated material.
\end{itemize}
In both cases the granulated material consists of $\sim$1 mm diameter frozen beads.
The cup choice changes the NMR filling factor and the degree of coupling to the resonant circuit. This becomes especially relevant for AFP in large, highly polarized samples where radiation damping and superradiant effects may occur \cite{Bloom1957,Odehnal1973,Reichertz1994}.

\subsection{Irradiated ammonia and deuterated ammonia}
\label{sec:irradiation}
The irradiated \ce{NH3} and \ce{ND3} used here were prepared by electron irradiation at the Medical Industrial Radiation Facility (MIRF) at the National Institute of Standards and Technology (NIST) in Gaithersburg, Maryland, using a traveling-wave linac and a \(\sim\)14~MeV electron beam at \(\sim\)10~\(\mu\)A to deliver an integrated fluence of order \(10^{17}~e^-/\mathrm{cm}^2\) \cite{Keller2013,Meyer2004}. In \ce{NH3}, warm (\(\sim\)87 K) irradiation produces \(\ce{NH2}\) paramagnetic centers \cite{Meyer2004,Keller2017} that serve as the electron-spin system required for DNP. For \ce{ND3}, the dominant free radical produced by warm irradiation is also believed to be \(\ce{NH2}\) \cite{Meyer1984ND3}, and only warm-irradiated material is used in the present work. By contrast, for cold (\(\sim\)1 K) irradiated \ce{ND3}, the relevant free radical and the DNP mechanism remain uncertain \cite{Keller2017}.

The irradiation was performed under a \(\sim\)87~K liquid-argon bath \cite{Keller2013} and the prepared material was stored in liquid nitrogen to preserve the paramagnetic complex. The nominal paramagnetic-center density for the irradiated materials is of order \(10^{19}\) spins/g.

\subsection{Butanol-based target materials}
In addition to irradiated ammonia, we studied both TEMPO-doped butanol and irradiated deuterated butanol. For the TEMPO-doped samples, the nominal radical density was \(2.2\times 10^{19}\) spins/g (from the known dopant loading). Proton (n-butanol) and deuteron (deuterated butanol) measurements used the same cryostat, target insert, and general AFP control scheme, but the RF/NMR circuits were tuned and matched separately at the relevant Larmor frequencies. Consequently, the proton and deuteron measurements should not be interpreted as having identical \(B_1\), resonant bandwidth, quality factor, or absolute adiabaticity. The fitted AFP sequence discussed in Sec.~\ref{sec:fit_results} corresponds to the irradiated deuterated butanol sample, but the same spin-1 lineshape analysis can be applied to other spin-1 materials with similar spectra.

\section{AFP theory: from rotating-frame adiabaticity to solid-state limits}
\label{sec:afp_theory}

\subsection{Rotating-frame description and Landau--Zener adiabaticity}
We summarize AFP in the standard rotating-frame picture \cite{Abragam1961,Bloch1946Induction}. For a spin system in a static field \(B_0\hat{z}\), driven by a transverse RF field \(B_1\) near resonance, the rotating-wave approximation yields an effective Hamiltonian
\begin{equation}
\mathcal{H}_{\mathrm{rot}}(t) = -\hbar \Delta\omega(t)\, I_z - \hbar \gamma B_1\, I_x,
\end{equation}
where \(\Delta\omega(t) = \omega(t) - \omega_0\) is the instantaneous detuning and \(\omega_0 = \gamma B_0\). The dynamics can be visualized as precession about an effective field
\begin{equation}
\mathbf{B}_\mathrm{eff}(t) = \left(B_1,\,0,\,\frac{\Delta\omega(t)}{\gamma}\right),
\end{equation}
which sweeps from \(+\hat{z}\) to \(-\hat{z}\) as \(\Delta\omega(t)\) crosses zero. In the ideal adiabatic limit, the macroscopic magnetization follows \(\mathbf{B}_\mathrm{eff}\), producing inversion.

For a linear sweep \(\Delta\omega(t)=\alpha t\), the two-level crossing problem yields the Landau--Zener transition probability \cite{Landau1932,Zener1932}
\begin{equation}
P_{\rm adiab} = \exp\!\left(-\frac{\pi (\gamma B_1)^2}{2|\alpha|}\right),
\end{equation}
so that inversion approaches unity when \((\gamma B_1)^2/|\alpha|\gg 1\). In practice for large solid targets, the spectrum is inhomogeneously broadened and the relevant condition must hold across the full line width and for the spatial distribution of \(B_1\).

\subsection{Spin-temperature (Provotorov) framework in solids}
In DNP target materials, nuclear spins form strongly interacting reservoirs and are coupled to paramagnetic centers. For broad lines, it is often useful to describe the nuclear system by spin temperatures (or inverse temperatures) associated with a Zeeman reservoir and a dipolar reservoir \cite{Provotorov1962,Goldman1968,Goldman1970}. In this picture, RF irradiation and frequency sweeps exchange energy between reservoirs; the AFP efficiency is then limited by:
\begin{itemize}
\item \textbf{Non-adiabaticity} (finite sweep rate or insufficient \(B_1\)),
\item \textbf{Relaxation in the rotating frame} (\(T_{1\rho}\)), and
\item \textbf{Spin diffusion and spectral diffusion} that redistribute populations across the broadened line.
\end{itemize}
Goldman \textit{et al.} derived quantitative expressions for ``single-shot'' passage in solids using the Provotorov equations, including the dependence on the adiabaticity parameter beyond only limiting cases \cite{Goldman1968}. Parfenov and Prudkoglyad used this framework to explain the poor AFP efficiency observed for proton frozen-spin target materials with high paramagnetic impurity densities \cite{Parfenov1984}.

A widely used engineering summary is the adiabaticity parameter
\begin{equation}
\begin{split}
A &\equiv \frac{\gamma B_1^2}{\mathrm{d}B/\mathrm{d}t}
\\
&\quad \text{(equivalently }
A = \frac{(\gamma B_1)^2}{\mathrm{d}\omega/\mathrm{d}t}
\\
&\quad \text{for a linear frequency sweep)}
\end{split}
\label{eq:A_def}
\end{equation}
and robust AFP typically requires \(A \gg 1\) across the line and across the sample \cite{Hautle1995}. When an independent absolute \(B_1\) calibration is not available but a set of measurements is made with the same coil and fixed sweep width, it is still useful to quote a relative adiabaticity scale,
\begin{equation}
A_{\rm rel}\equiv \frac{A}{A_0}
=\left(\frac{V_{\rm RF}}{V_0}\right)^2
 \left(\frac{\tau}{\tau_0}\right),
\label{eq:Arel_def}
\end{equation}
where \(V_{\rm RF}\) is the applied RF drive setting, \(\tau\) is the sweep time, and \(A_0\) is a reference condition measured with the same sweep width and coil. This relative scale is not a substitute for a calibrated \(B_1\), but it combines the experimentally varied sweep time and RF drive into the parameter most directly associated with adiabaticity.

\subsection{Radiation damping, auto-oscillation, and superradiance}
\label{sec:rad_damp}
In large, highly polarized samples coupled to a high-Q resonant circuit, the nuclear magnetization can significantly drive the circuit. The resulting radiation damping can break the symmetry between AFP starting from \(+P\) and \(-P\) \cite{Bloom1957,Odehnal1973}. When the effective negative resistance produced by the spin system compensates circuit losses, auto-oscillation and superradiant emission can occur, leading to strongly asymmetric reversal efficiencies \cite{Reichertz1994}. This phenomenon is particularly relevant for large \ce{NH3} samples at 5~T/1~K, where the proton frequency is high and RF coils are often operated with high Q.

These effects motivate two practical choices in our measurements:
(i) using a small 1~g cup to reduce coupling and suppress superradiance during precision studies, and
(ii) explicitly studying the dependence of AFP efficiency on the initial polarization in a large 7~g \ce{NH3} configuration (Sec.~\ref{sec:nh3_dep}).

\subsection{Spin-1 (deuteron) AFP with quadrupolar splitting}
\label{sec:spin1_afp}
For deuterons (\(I=1\)) the NMR spectrum is split by quadrupolar interactions, producing two overlapping \(\Delta m=\pm 1\) transitions. In polycrystalline materials this yields the Pake doublet \cite{Pake1948,Dulya1997,Keller2020}. AFP in such a system is not a single isolated two-level crossing: inversion proceeds through sequential passages across the two transitions. As shown experimentally by Hautle \textit{et al.}, complete inversion of the vector polarization can require a full frequency sweep followed by an additional half-sweep under typical conditions \cite{Hautle1992}.

This multilevel structure is also what enables tensor polarimetry: the populations \(n_{+1}, n_0, n_{-1}\) can be driven into non-Boltzmann distributions during interrupted or partial sweeps, producing manipulated states with nontrivial \(P\) and \(Q\) that must be extracted from lineshape information rather than a single equilibrium intensity ratio \cite{Keller2020}.

\section{Spin-1 lineshape and vector/tensor polarimetry}
\label{sec:afp_transitions}

\subsection{Deuteron lineshape}
The deuteron, a spin-1 nucleus, experiences both Zeeman and quadrupolar interactions in the presence of an external magnetic field. The magnetic sublevels \(\ket{m}\) (\(m=+1,0,-1\)) are split by the combined interactions. The corresponding energy levels can be expressed as~\cite{Abragam1961,Dulya1997}
\begin{equation}
\begin{aligned}
E_m &= -\hbar \omega_d m \\
    &\quad + \hbar \omega_q
      \left[3 \cos^2 \theta - 1 + \eta \sin^2 \theta \cos 2\phi \right]
      (3m^2 - 2),
\end{aligned}
\end{equation}
where \(\omega_d\) and \(\omega_q\) are the deuteron Larmor and quadrupolar frequencies. The observable \(\Delta m=\pm1\) transitions occur at
\begin{align}
\omega_+ &= \omega_d + 3\omega_q (3 \cos^2 \theta - 1), \\
\omega_- &= \omega_d - 3\omega_q (3 \cos^2 \theta - 1),
\end{align}
producing the characteristic Pake doublet \cite{Pake1948}. The two absorption components, denoted \(I_+(\omega)\) and \(I_-(\omega)\), overlap in a polycrystalline sample and must be separated by a lineshape model before vector and tensor polarizations can be inferred.

In practice we use the Dulya representation of the Pake lineshape~\cite{Dulya1997}. We define the dimensionless normalized frequency variable
\begin{equation}
R = \frac{\nu - \nu_d}{3\nu_Q},
\end{equation}
where \(\nu\) is the measured frequency, \(\nu_d\) is the deuteron Larmor frequency, and \(\nu_Q\) is the quadrupolar frequency scale. In terms of \(R\), the line shape for each transition is
\begin{equation}
\scalebox{0.85}{$\displaystyle
\begin{aligned}
f(R,\epsilon) = {}&\frac{1}{2\pi X}\Bigg[
  2\cos\!\left(\tfrac{\alpha}{2}\right)
   \left(
      \arctan\!\left(
         \frac{Y^{2}-X^{2}}{2YX\sin(\alpha/2)}
      \right)
      + \frac{\pi}{2}
   \right) \\
  &+ \sin\!\left(\tfrac{\alpha}{2}\right)
   \ln\!\left(
      \frac{Y^{2}+X^{2}+2YX\cos(\alpha/2)}
           {Y^{2}+X^{2}-2YX\cos(\alpha/2)}
   \right)
\Bigg],
\end{aligned}
$}
\label{eq:dulya}
\end{equation}
where
\begin{align}
X^2 &= \sqrt{\Delta^2 + (1 - \epsilon R - \eta\cos(2\phi))^2}, \\
Y   &= \sqrt{3 - \eta\cos(2\phi)}, \\
\cos\alpha &= \frac{1 - \epsilon R - \eta\cos(2\phi)}{X^2}.
\end{align}
Here \(\Delta\) denotes the Lorentzian dipolar width, \(\eta\) is the asymmetry parameter, and \(\phi\) is the azimuthal angle of the molecular axis relative to the magnetic field \cite{Abragam1961,Cohen1957}. The parameter \(\epsilon=\pm1\) specifies the orientation of the two Pake components.

\begin{figure}[h!]
\centering
\includegraphics[height=67mm]{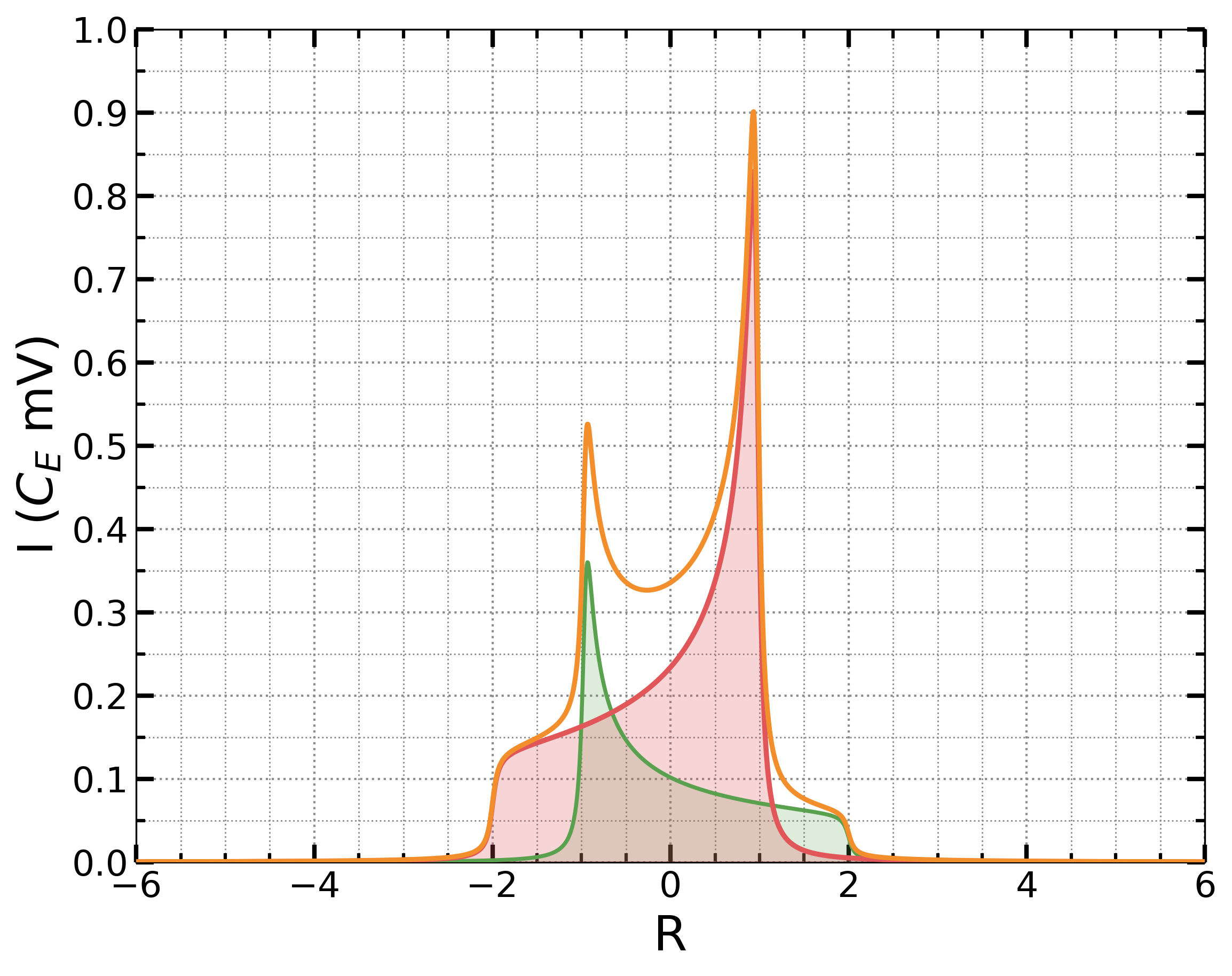}
\caption{Example NMR lineshape of a spin-1 target with non-cubic symmetry, showing the two overlapping absorption components \(I_{+}\) and \(I_{-}\). The measured NMR signal is the sum of these components.}
\label{pake}
\end{figure}

\subsection{Vector and tensor polarization definitions}
\label{sec:polarimetry}
For spin-1 nuclei, we define the vector polarization \(P\) and tensor polarization \(Q\) in terms of the Zeeman sublevel populations \cite{Dulya1997,Keller2020}
\begin{align}
P &= n_{+1} - n_{-1},\\
Q &= n_{+1} + n_{-1} - 2n_0,
\end{align}
with \(n_{+1}+n_0+n_{-1}=1\). Inverting these relations gives
\begin{align}
n_{+1} &= \frac{1}{3}  + \frac{1}{2} P + \frac{1}{6} Q, \label{eq:nplus} \\
n_{0} &= \frac{1 - Q}{3},        \label{eq:n0} \\
n_{-1} &= \frac{1}{3}  - \frac{1}{2} P + \frac{1}{6} Q.  \label{eq:nminus}
\end{align}
The two absorption lines correspond to population differences,
\begin{align}
I_{+}(\omega) &\propto (n_{+1}-n_0)\, f_{+}(\omega), \label{eq:Iplus_def}\\
I_{-}(\omega) &\propto (n_{0}-n_{-1})\, f_{-}(\omega), \label{eq:Iminus_def}
\end{align}
where \(f_{\pm}\) are the common powder-pattern line shapes broadened by the instrumental response and inhomogeneity.

Under global thermal-equilibrium conditions, the system satisfies Boltzmann statistics and the integrated intensity ratio \(r=I_+/I_-\) uniquely determines both polarizations~\cite{Dulya1997,Keller2017}:
\begin{align}
P &= \frac{r^2 - 1}{r^2 + r + 1}, \label{eq:Pn_from_r} \\
Q &= \frac{r^2 - 2r + 1}{r^2 + r + 1}. \label{eq:Qn_from_r}
\end{align}
During AFP, especially in interrupted or partial sweeps, this global Boltzmann relation is generally not valid. We use the term ``half-flip'' for such manipulated spectra when the AFP sequence has changed only part of the spin-1 population distribution or has not yet produced a complete inversion. In these cases the polarizations are obtained from the fitted component areas,
\begin{align}
P &= C\,(I^+ + I^-), \label{eq:Pn_from_areas} \\
Q &= C\,(I^+ - I^-), \label{eq:Qn_from_areas}
\end{align}
where \(C\) is a calibration constant determined from thermal-equilibrium spectra. These area relations can be applied to AFP-manipulated spectra only if \(I^+\) and \(I^-\) are extracted using a physically constrained decomposition of the two overlapping components.

\subsection{Joint manipulated-lineshape analysis}
\label{sec:new_method}
The analysis of AFP-manipulated NMR signals relies on three principles established for selective semi-saturating RF manipulation in spin-1 systems~\cite{Keller2017,Clement:2023eun}. First, differential binning partitions the frequency domain so that each bin can be treated as a local subset of the total population. Second, the rates response quantifies how RF irradiation in one frequency region reduces the area of the affected absorption line while producing a compensating enhancement of the opposing line with one-half the depleted area. Third, spin-temperature consistency requires that overlapping absorption lines at a fixed frequency maintain an intensity ratio corresponding to a unique local effective spin temperature, even when the system is not in global thermal equilibrium.

For a background-subtracted spectrum, the model signal is written as
\begin{equation}
S(\omega)=\mathcal{N}\left[S_{+}(\omega)+S_{-}(\omega)\right],
\label{eq:joint_model_total}
\end{equation}
with
\begin{align}
S_{+}(\omega) &= \sum_j a_{+,j}\,\Pi_j(\omega)\,f_{+}(\omega;\bm{\lambda}), \\
S_{-}(\omega) &= \sum_j a_{-,j}\,\Pi_j(\omega)\,f_{-}(\omega;\bm{\lambda}),
\label{eq:joint_model_components}
\end{align}
where \(f_{\pm}\) are the equilibrium Pake/Dulya component shapes, \(\bm{\lambda}\) denotes shared line-center, quadrupole-splitting, and broadening parameters, \(\Pi_j\) is a piecewise binning of the frequency coordinate, and \(a_{\pm,j}\) are local transition amplitudes. The overall normalization is denoted \(\mathcal{N}\) to avoid confusion with the adiabaticity parameter \(A\).

The amplitudes \(a_{+,j}\) and \(a_{-,j}\) are constrained within each bin by a local spin-temperature-consistency condition,
\begin{equation}
\frac{a_{+,j}}{a_{-,j}} = r_s(R_j,T_{s,j}),
\label{eq:joint_model_rs}
\end{equation}
where \(r_s(R_j,T_{s,j})\) is the equilibrium asymmetry ratio at coordinate \(R_j\) for a local effective spin temperature \(T_{s,j}\) \cite{Clement:2023eun}. Hole burning and AFP distortion are introduced through a coupled response of mirrored bins,
\begin{equation}
\Delta a_{\mp,j'} = \frac{1}{2}\,\Delta a_{\pm,j},
\qquad R_{j'}\simeq -R_j,
\label{eq:joint_model_rates}
\end{equation}
after convolution with a finite-width RF-response kernel that accounts for field inhomogeneity, circuit bandwidth, and the nonzero spectral width of the manipulation. For AFP data this kernel may depend on sweep direction and sweep rate, allowing the same framework to describe localized hole burning and sweep-history-dependent AFP distortion.

The complete model is fit to the digitized spectrum with weighted nonlinear least squares, using a constant point-by-point uncertainty determined from the root-mean-square deviation (RMSD) of the signal tails as described in Appendix~\ref{subsec:errors}. The fitted component areas \(I^{\pm}=\int S_{\pm}(\omega)\,d\omega\) are then converted to \(P\) and \(Q\) through Eqs.~(\ref{eq:Pn_from_areas}) and (\ref{eq:Qn_from_areas}), with statistical uncertainties propagated from the full covariance matrix of the joint fit.

\section{AFP experimental parameters and optimization}
\label{sec:afp_optimization}

AFP performance is governed primarily by the sweep rate and transverse RF field through the adiabaticity parameter defined in Eq.~(\ref{eq:A_def}). For a linear frequency sweep of width \(\Delta f\) and duration \(\tau\),
\begin{equation}
\frac{\mathrm{d}\omega}{\mathrm{d}t}=2\pi\frac{\Delta f}{\tau},
\qquad
B_1=\sqrt{\frac{A}{\gamma^2}\frac{\mathrm{d}\omega}{\mathrm{d}t}}.
\label{eq:B1_req}
\end{equation}
Thus increasing \(B_1\) or lengthening the sweep increases \(A\), whereas increasing the sweep width at fixed \(\tau\) decreases it. The same RF drive setting does not imply the same \(B_1\) for protons and deuterons, because the resonance frequency, coil impedance, matching network, and quality factor differ substantially between \(213~\mathrm{MHz}\) and \(32.7~\mathrm{MHz}\).

For the deuteron AFP conditions used in the lineshape analysis, representative parameters are listed in Table~\ref{tab:afp_params}. The table should be understood as an operating scale for this configuration rather than a universal prescription. The circuit was tuned to cover the AFP sweep bandwidth while keeping reflected power and cryogenic heat load acceptable.

\begin{table}[t]
\centering
\caption{Representative deuteron AFP operating parameters and RF-chain settings used for the manipulated-lineshape measurements.}
\label{tab:afp_params}
\setlength{\tabcolsep}{4pt}
\renewcommand{\arraystretch}{1.05}
\begin{tabularx}{\columnwidth}{@{}lX@{}}
\toprule
\textbf{Quantity} & \textbf{Value} \\
\midrule
Static magnetic field & \(B_0 = 5~\mathrm{T}\) \\
Temperature & \(T \simeq 1~\mathrm{K}\) \\
Nucleus & \(^2\)H (deuteron) \\
Gyromagnetic ratio & \(\gamma_d/2\pi \approx 6.535~\mathrm{MHz/T}\) \\
Central Larmor frequency & \(f_0 = 32.7~\mathrm{MHz}\) \\
Sweep width / duration & \(\Delta f = 400~\mathrm{kHz}\), \(\tau = 0.2~\mathrm{s}\) \\
Angular sweep rate & \(\mathrm{d}\omega/\mathrm{d}t \approx 1.26\times10^{7}~\mathrm{rad/s^2}\) \\
Adiabaticity target & \(A \gtrsim 10\)--\(30\) \\
Estimated transverse field & \(B_1 \sim 2\)--\(5~\mathrm{G}\) \\
RF chain & R\&S generator \(\rightarrow\) LZY-1X+ amplifier \(\rightarrow\) directional coupler \(\rightarrow\) coax \(\rightarrow\) coil \\
Bandwidth / matching & \(\Delta f_{\mathrm{BW}}\simeq f_0/Q \gtrsim \Delta f\), requiring \(Q\lesssim 80\) \\
\bottomrule
\end{tabularx}
\end{table}

The proton \ce{NH3} efficiency study discussed below did not include an independent absolute \(B_1\) calibration from slightly saturating passages. To avoid overinterpreting those measurements, Table~\ref{tab:nh3_init} reports a relative adiabaticity parameter \(A_{\rm rel}\) computed from Eq.~(\ref{eq:Arel_def}) using the fixed 400~kHz sweep width and the RF drive setting as a proxy for \(B_1\). Future proton AFP measurements should include a direct \(B_1\) calibration so that the efficiency can be mapped on an absolute \(A\) scale.

\section{Experimental AFP efficiency results}
\label{sec:results}

\subsection{AFP efficiencies in multiple materials}
Table~\ref{tab:materials} summarizes the AFP efficiencies measured in this work for several materials relevant to polarized targets. We define the AFP efficiency as
\begin{equation}
\epsilon \equiv \left|\frac{P_f}{P_i}\right|,
\end{equation}
where \(P\) denotes the relevant polarization observable: vector polarization for spin-1 and ordinary polarization for spin-\(\tfrac{1}{2}\).

\begin{table*}[t]
\centering
\small
\setlength{\tabcolsep}{4pt}
\renewcommand{\arraystretch}{1.05}
\caption{Summary of AFP efficiencies measured in this work. The \ce{NH3} entry was acquired in the large \((\sim 7~\mathrm{g})\) configuration and exhibits direction- and setup-dependent behavior; details are given in Table~\ref{tab:nh3_init}.}
\label{tab:materials}
\begin{tabularx}{\textwidth}{@{}>{\raggedright\arraybackslash}X l l l l l l l@{}}
\toprule
Material 
& Nucleus 
& Paramagnetic centers 
& Spins/g 
& \(\epsilon\) 
& \(\delta\epsilon\) 
& Cup 
& \(P_i\)(\%) \\
\midrule
Deuterated butanol + TEMPO 
  & \(^2\)H 
  & TEMPO 
  & \(2.2\times10^{19}\) 
  & 0.77 
  & 0.03 
  & 7 g 
  & 0.35 \\
n-butanol + TEMPO 
  & \(^1\)H 
  & TEMPO 
  & \(2.2\times10^{19}\) 
  & 0.38 
  & 0.04 
  & 1 g 
  & 0.82 \\
Irradiated deuterated butanol 
  & \(^2\)H 
  & Irradiation 
  & warm dose
  & 0.80 
  & 0.03 
  & 7 g 
  & 0.36 \\ 
Irradiated \ce{^{14}ND3} 
  & \(^2\)H 
  & Irradiation 
  & warm dose
  & 0.88 
  & 0.02 
  & 1 g 
  & 0.22 \\
Irradiated \ce{^{15}NH3} 
  & \(^1\)H 
  & Irradiation 
  & warm dose 
  & 0.48 
  & 0.04 
  & 7 g 
  & 0.92 \\
\bottomrule
\end{tabularx}
\end{table*}

\subsection{Efficiency dependence on initial NMR area in a 7 g \ce{NH3} target}
\label{sec:nh3_dep}
A dedicated study was performed using a large \((\sim 7~\mathrm{g})\) irradiated \ce{NH3} target at 5~T. The proton resonance was centered near 213~MHz with a 400~kHz sweep width. Table~\ref{tab:nh3_init} reports the measured NMR areas before and after AFP and the resulting efficiencies. Because the measurements used several sweep times and RF drive settings, the table also lists the relative adiabaticity parameter
\begin{equation}
A_{\rm rel}=\left(\frac{V_{\rm RF}}{100~\mathrm{mV}}\right)^2
\left(\frac{\tau}{0.25~\mathrm{s}}\right),
\end{equation}
which follows Eq.~(\ref{eq:Arel_def}) for the fixed 400~kHz sweep width. This quantity combines the two varied AFP settings into a single relative scale. It should not be interpreted as an absolute adiabaticity because the corresponding \(B_1\) was not independently calibrated.

For compactness, we report a normalized initial-polarization proxy \(|I_i|/I_{\max}\), where \(I_i\) is the measured initial NMR area and \(I_{\max}\) is the largest magnitude observed in this dataset. Absolute polarization conversion follows standard thermal-equilibrium calibration procedures \cite{Keller2013}.

\begin{table*}[t]
\centering
\small
\caption{AFP study in a large \((\sim 7~\mathrm{g})\) irradiated \ce{NH3} target at 5~T (proton frequency \(\approx 213~\mathrm{MHz}\), sweep width 400~kHz). \(A_{\rm rel}\) is the relative adiabaticity parameter normalized to the 100~mV, 0.25~s setting. \(I_i\) and \(I_f\) are the NMR areas before and after AFP. The last column flags entries where the final NMR area did not change sign.}
\label{tab:nh3_init}
\begin{tabular}{@{}cccccccccc@{}}
\toprule
\(\tau\) (s) & RF amp. & \(A_{\rm rel}\) & \(I_i\) & \(I_f\) & \(|I_i|/I_{\max}\) & \(\epsilon\) & \(\delta\epsilon\) & Direction & Flip? \\
\midrule
0.25 & 100 mV & 1.0  &  1.250 & -0.307 & 0.233 & 0.246 & 0.03 & \(+\rightarrow-\) & yes \\
0.25 & 600 mV & 36.0 &  1.130 &  0.562 & 0.210 & 0.497 & 0.06 & \(+\rightarrow-\) & partial \\
0.25 & 200 mV & 4.0  &  1.110 & -0.573 & 0.207 & 0.516 & 0.04 & \(+\rightarrow-\) & yes \\
0.20 & 200 mV & 3.2  &  0.945 & -0.522 & 0.176 & 0.552 & 0.04 & \(+\rightarrow-\) & yes \\
0.10 & 200 mV & 1.6  &  0.707 & -0.517 & 0.132 & 0.731 & 0.03 & \(+\rightarrow-\) & yes \\
0.10 & 600 mV & 14.4 &  0.666 & -0.374 & 0.124 & 0.562 & 0.04 & \(+\rightarrow-\) & yes \\
\midrule
0.40 & 200 mV & 6.4  & -5.370 &  0.707 & 1.000 & 0.132 & 0.02 & \(-\rightarrow+\) & yes \\
0.25 & 200 mV & 4.0  & -4.580 &  1.231 & 0.853 & 0.269 & 0.03 & \(-\rightarrow+\) & yes \\
0.25 & 600 mV & 36.0 & -4.480 &  1.130 & 0.834 & 0.252 & 0.03 & \(-\rightarrow+\) & yes \\
0.25 & 400 mV & 16.0 & -4.310 &  1.289 & 0.803 & 0.299 & 0.03 & \(-\rightarrow+\) & yes \\
0.10 & 600 mV & 14.4 & -4.280 &  0.678 & 0.797 & 0.158 & 0.02 & \(-\rightarrow+\) & yes \\
0.20 & 200 mV & 3.2  & -4.210 &  0.949 & 0.784 & 0.225 & 0.03 & \(-\rightarrow+\) & yes \\
0.10 & 200 mV & 1.6  & -0.781 &  0.125 & 0.145 & 0.160 & 0.03 & \(-\rightarrow+\) & yes \\
0.10 & 600 mV & 14.4 & -0.381 & -0.045 & 0.071 & 0.118 & 0.05 & \(-\rightarrow+\) & partial \\
\bottomrule
\end{tabular}
\end{table*}

The purpose of Table~\ref{tab:nh3_init} is to document the measured response of this large-sample configuration, not to establish a universal polarization-scaling law for AFP. The positive-polarization entries show higher measured efficiencies at lower initial NMR area, but they were not all acquired at a fixed \(A_{\rm rel}\), and the absolute \(B_1\) scale was not calibrated. The negative-polarization branch shows a pronounced direction-dependent response at large \(|I_i|\). These observations were reproducible in this apparatus, but we interpret them conservatively: they likely reflect sample--circuit back-action, radiation damping or superradiant behavior, or another coil/material-specific mechanism rather than a simple intrinsic increase of AFP efficiency at low initial polarization.

\section{Manipulated spin-1 lineshape analysis}
\label{sec:fit_results}

\subsection{AFP-manipulated deuteron lineshape fits}
\begin{figure}[]
  \centering

  \begin{subfigure}[t]{0.8\columnwidth}
    \centering
    \includegraphics[width=\linewidth]{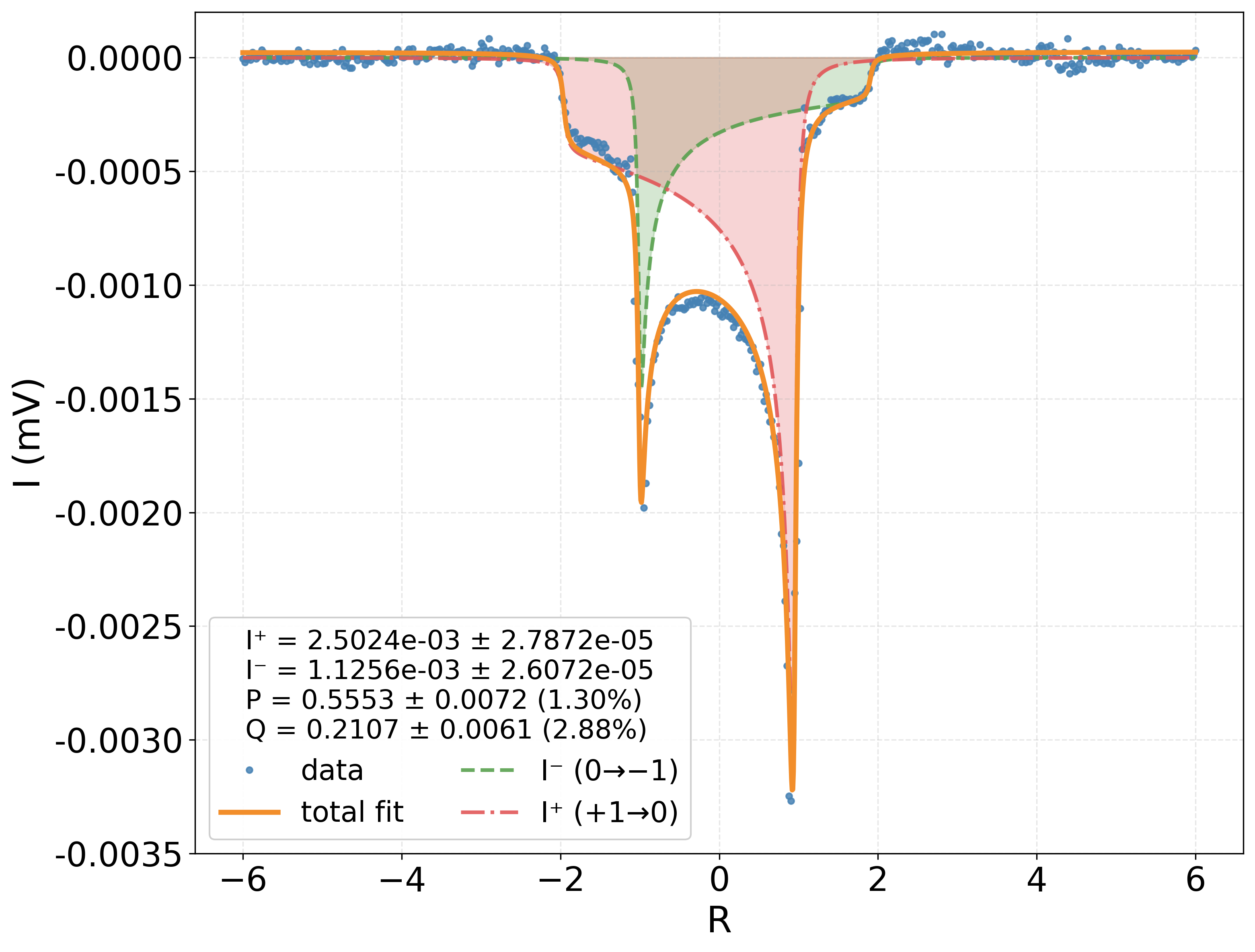}
    \caption{Initial equilibrium signal from irradiated deuterated butanol (before AFP).}
    \label{fig:afp_initial}
  \end{subfigure}

  \vspace{0.6em}

  \begin{subfigure}[t]{0.8\columnwidth}
    \centering
    \includegraphics[width=\linewidth]{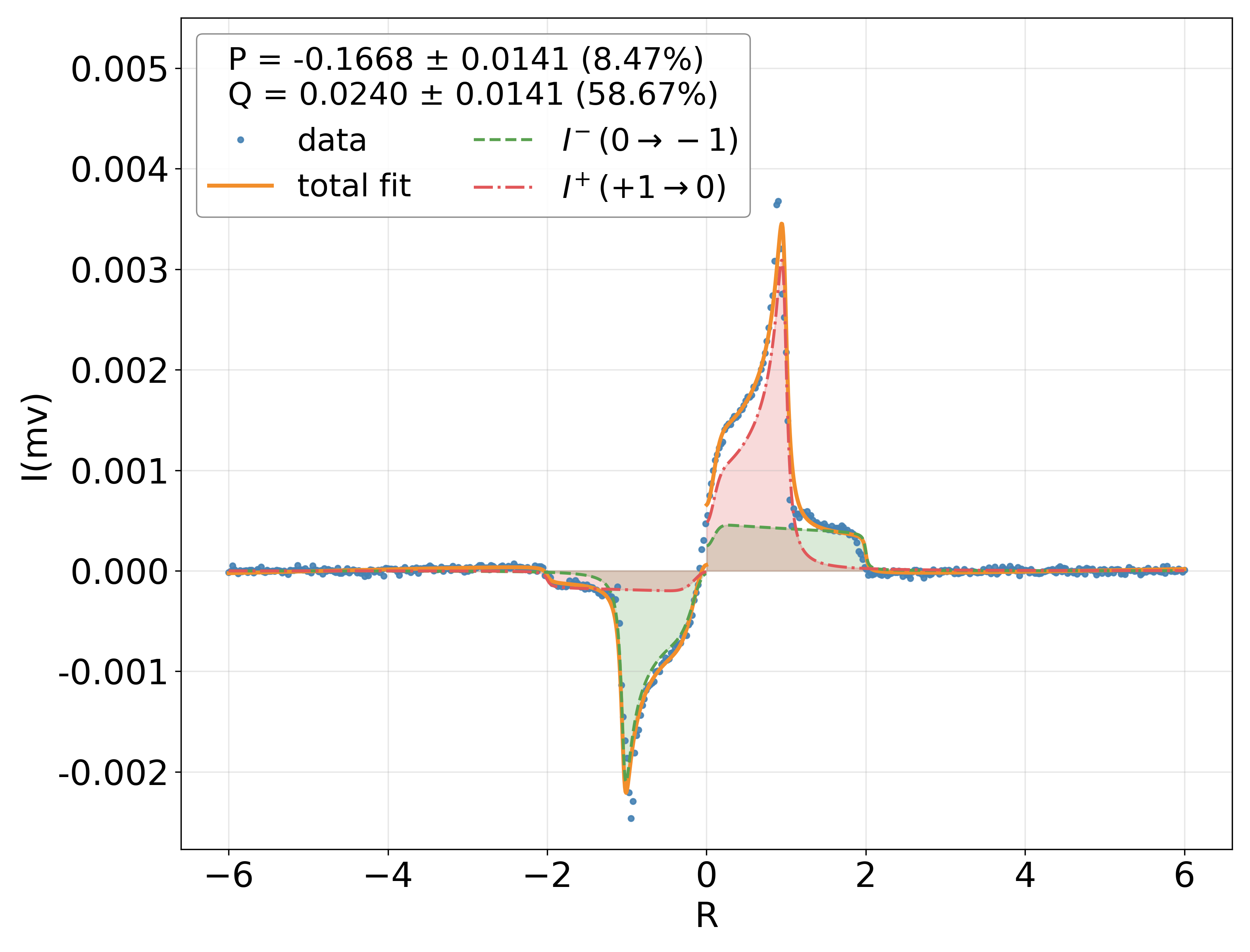}
    \caption{Intermediate AFP state in irradiated deuterated butanol with localized hole burning.}
    \label{fig:afp_partial}
  \end{subfigure}

  \vspace{0.6em}

  \begin{subfigure}[t]{0.8\columnwidth}
    \centering
    \includegraphics[width=\linewidth]{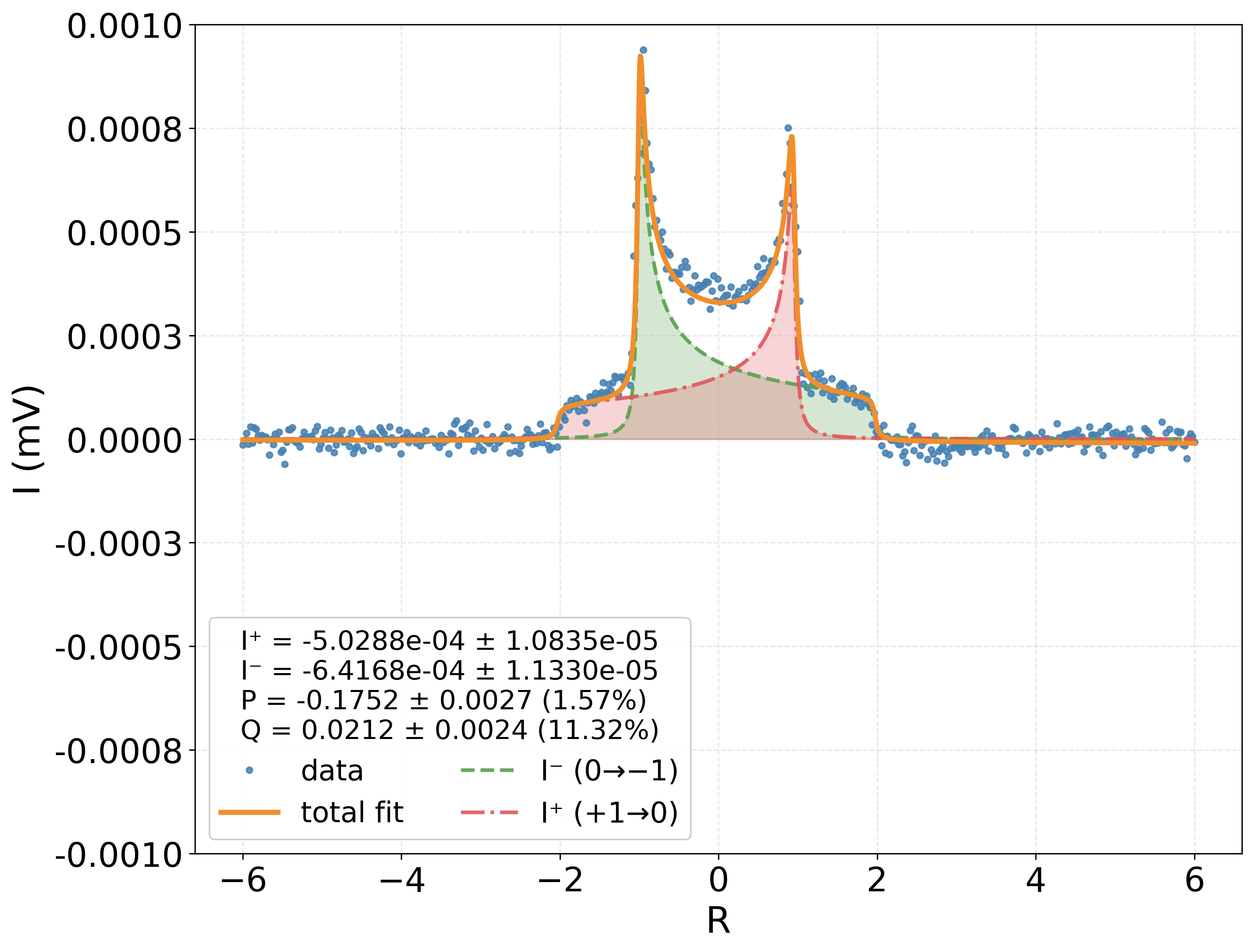}
    \caption{Final irradiated deuterated butanol state after complete AFP inversion.}
    \label{fig:afp_complete}
  \end{subfigure}

  \caption{Evolution of the deuteron NMR signal in irradiated deuterated butanol during an AFP sweep: (a) initial equilibrium configuration, (b) intermediate AFP state exhibiting hole burning, and (c) fully inverted state. The lower panel in each subfigure shows the fit residuals.}
  \label{fig:afp_sequence}
\end{figure}

During an AFP sweep, the applied RF field can produce localized saturation at the swept frequencies, leading to hole burning in the overlapping spin-1 absorption lines. As established in the rates-response formalism (Sec.~IV of Ref.~\cite{Clement:2023eun}), depletion of one transition within a narrow frequency window is accompanied by a compensating enhancement of the opposing transition with one-half the depleted area at the mirrored coordinate \(R\rightarrow -R\). These effects are undesirable for maximizing AFP efficiency, but they provide stringent tests of the manipulated-lineshape fitting technique. The fitted spectra in Fig.~\ref{fig:afp_sequence} are taken from an irradiated deuterated butanol AFP sequence and illustrate the spectral evolution over successive passages. The curves shown are joint manipulated-lineshape fits using the common Dulya/Pake kernel and the coupled-response constraints of Sec.~\ref{sec:new_method}.

Figure~\ref{fig:afp_initial} shows the deuteron line after DNP but before AFP, where the calibration constant \(C\) and baseline polarizations are established. Figure~\ref{fig:afp_partial} shows an intermediate state after a full low-to-high frequency sweep, where the local depletion/gain pattern expected from the rates response is evident. Figure~\ref{fig:afp_complete} shows the final state following the subsequent half-sweep. In all three cases, the joint manipulated-lineshape model describes the data adequately throughout the AFP sequence.

\begin{figure}[h]
  \centering
  \includegraphics[width=\columnwidth]{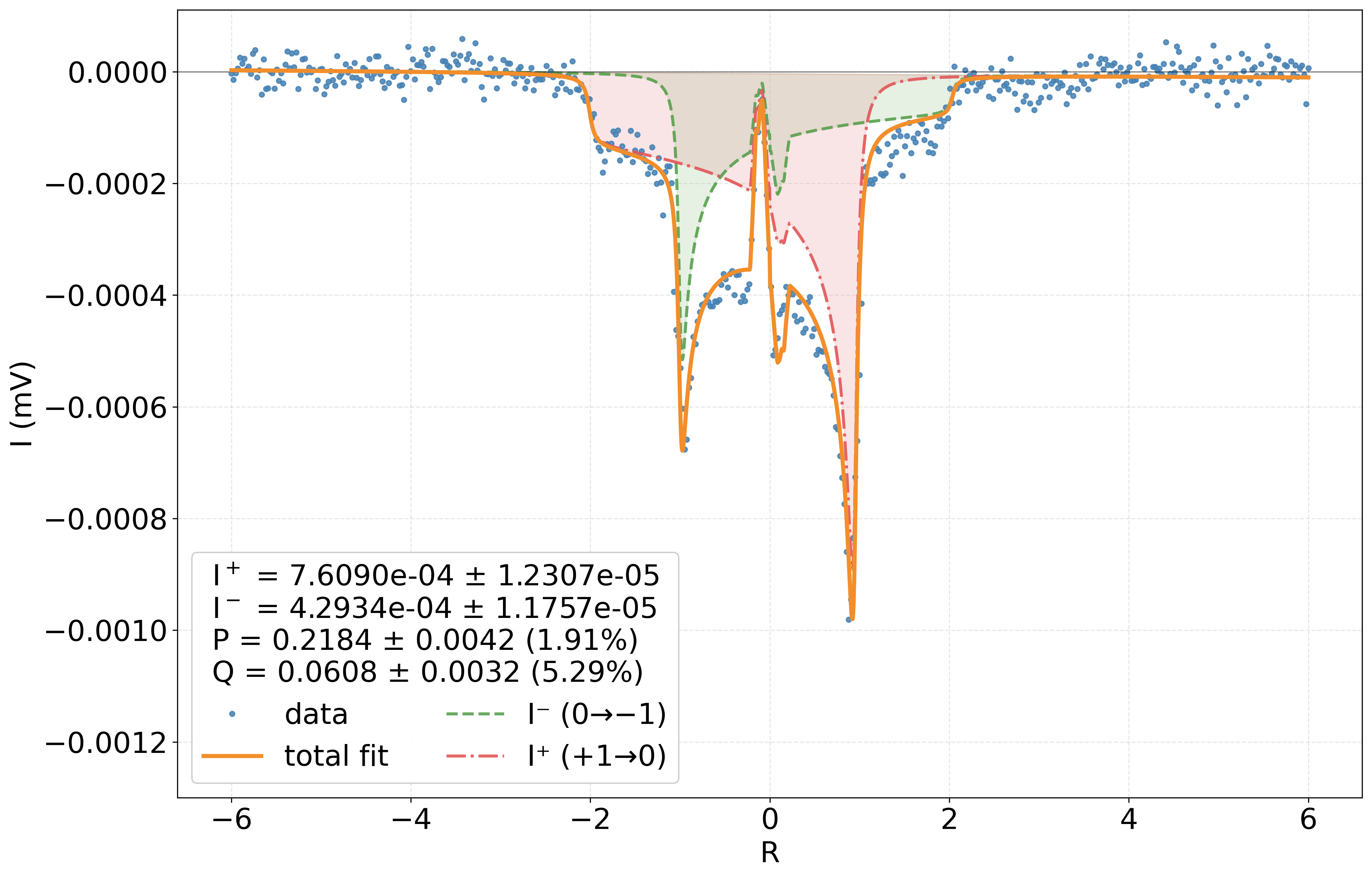}\par\vspace{0.6em}
  \caption{AFP-induced hole burning in irradiated deuterated butanol. The dip near \(R\lesssim0\) is an unintended burn produced by continued irradiation after the sweep terminated.}
  \label{fig:afp_experimental}
\end{figure}

Figure~\ref{fig:afp_experimental} demonstrates hole burning caused by an RF-generator spur after a single low-to-mid frequency AFP sweep. The depleted region near the center of the \(R<0\) transition is accompanied by a corresponding enhancement in the opposite transition, in agreement with the rates-response picture. The narrow dip observed near \(R\lesssim0\) arises from continued irradiation by the RF generator at a fixed frequency after the sweep terminated, producing an unintended localized burn. Similar generator-related effects have been reported previously \cite{Hautle1992,Hautle1995}; in practice, they can be avoided by reversing the sweep direction or using generators with faster response.

\begin{figure}[htbp]
  \centering
  \includegraphics[width=\columnwidth]{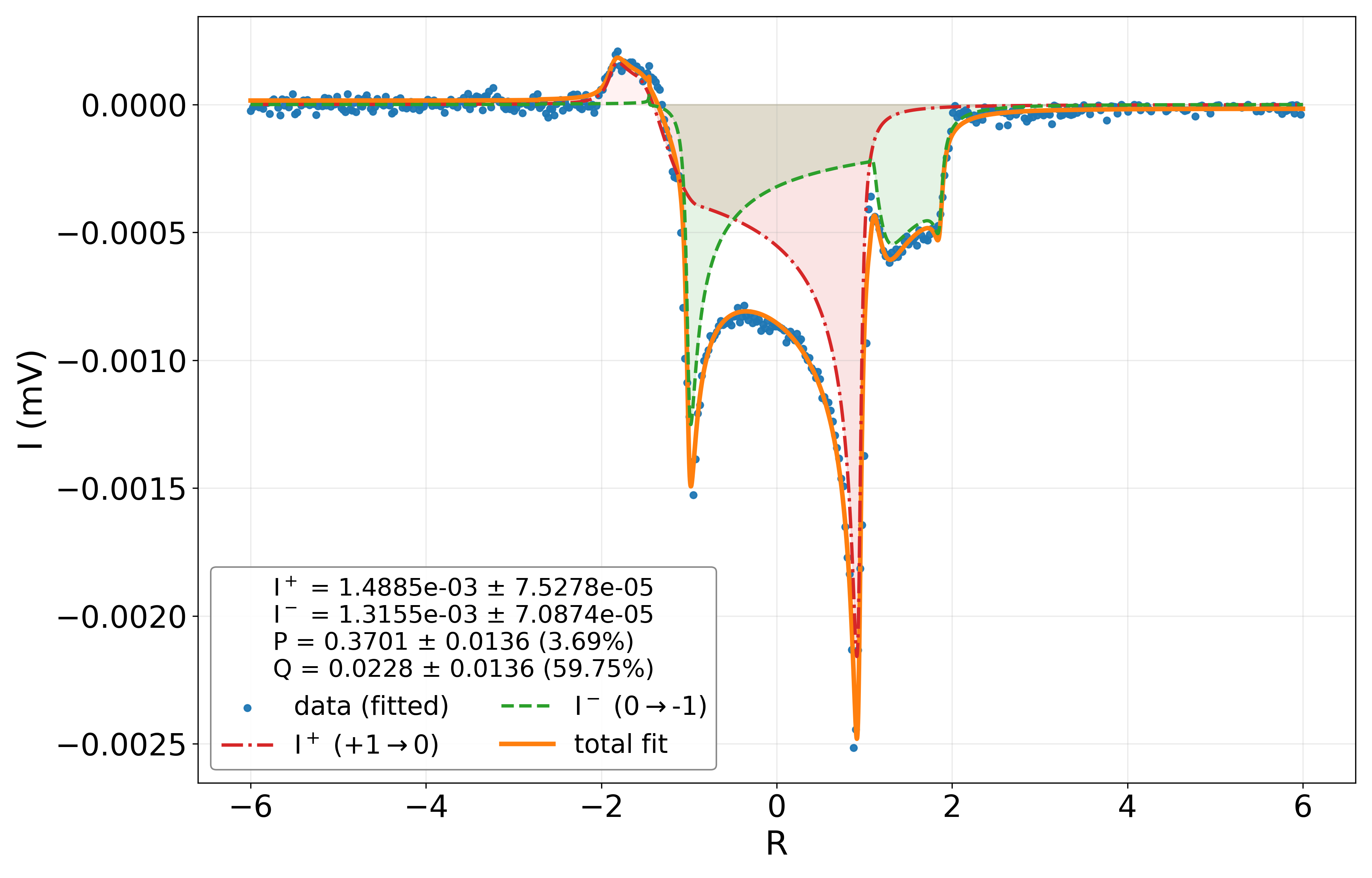}\par\vspace{0.6em}
  \caption{Pedestal AFP manipulation in irradiated deuterated butanol, with additional hole burning on the left pedestal. The burn reduces \(I_+\) in the shoulder region while enhancing \(I_-\) in the right pedestal at half the depletion rate, demonstrating the rates-response mechanism.}
  \label{fig:pedestal_burning}
\end{figure}

Pedestal flipping with AFP provides another example of a highly manipulated spin-1 line. By selectively flipping one pedestal, the AFP sequence can modify \(I^+\) or \(I^-\) and therefore change \(Q=C(I^+-I^-)\) \cite{Keller2020}. Figure~\ref{fig:pedestal_burning} demonstrates a pedestal flip with additional hole burning on the left pedestal; this dataset is used as a fitting example for tensor-polarization depletion.

\subsection{Spin-1 half-flip polarimetry demonstration in irradiated deuterated butanol}
Using the spin-1 analysis method of Sec.~\ref{sec:new_method}, we extract \(P\) and \(Q\) not only before and after AFP but also during interrupted or half-sweep states in the irradiated deuterated butanol sequence shown in Fig.~\ref{fig:afp_sequence}. In these states the spectrum typically violates the Boltzmann constraint \(Q(P)\) expected for a single spin temperature \cite{Keller2020}. By fitting the two overlapping transition components jointly, with shared lineshape parameters and local rates-response and spin-temperature-consistency constraints, we determine the instantaneous sublevel populations and thereby obtain a consistent polarimetry even in non-equilibrium AFP states.

\section{Discussion}
The material survey (Table~\ref{tab:materials}) is consistent with the long-standing observation that deuteron-bearing materials can achieve higher AFP efficiencies than proton materials under suitable target and circuit conditions \cite{Hautle1992,Hautle1995}. The large-sample \ce{NH3} results in Table~\ref{tab:nh3_init} should be interpreted more cautiously. They show a reproducible dependence of the measured reversal efficiency on initial NMR area, reversal direction, and the relative adiabaticity scale \(A_{\rm rel}\), but they do not by themselves establish an intrinsic polarization dependence of AFP efficiency.

This distinction is important because conventional rotating-frame relaxation arguments do not imply that lower polarization should generally improve AFP efficiency. The observed \ce{NH3} behavior is therefore more plausibly associated with the specific large-sample coil/material configuration, for example sample--circuit back-action, radiation damping, superradiant behavior, or another mechanism not isolated in the present measurements. Since the proton \(B_1\) was not absolutely calibrated, these data are best viewed as an empirical map of this apparatus rather than a universal optimization curve. A controlled measurement at fixed calibrated \(A\), with direct measurements of circuit \(Q\), \(B_1\), and induced-field thresholds, is needed to separate ordinary AFP adiabaticity from circuit-mediated effects.

The manipulated-lineshape analysis provides a separate methodological result. It allows \(P\) and \(Q\) to be extracted from spin-1 spectra that are distorted by AFP, hole burning, or interrupted sweeps, without imposing a single global spin temperature. This is useful for experiments that employ multi-step AFP sequences, full-plus-half sweeps in spin-1 targets, or combinations of AFP with hole burning and selective RF manipulation \cite{Keller2026}. Although the demonstrated sequence uses irradiated deuterated butanol, the same formalism is applicable to other manipulated spin-1 spectra, including \ce{ND3}, when the relevant lineshape parameters are known.

\section{Conclusions}
We have presented AFP measurements and manipulated-lineshape analysis in a 5~T/1~K polarized-target system with interchangeable 1~g and 7~g target cups. The survey of materials studied here shows that AFP performance remains strongly material dependent: the deuteron-bearing systems \ce{ND3} and deuterated butanol achieve higher measured efficiencies than the proton systems under the tested conditions. For the large \ce{NH3} sample, the measured reversal efficiency varies reproducibly with initial NMR area, sweep direction, and relative adiabaticity. We do not interpret this as evidence that low initial polarization intrinsically increases AFP efficiency. Rather, the data document the behavior of a specific large-sample, high-frequency sample--coil configuration in which circuit-mediated effects such as radiation damping or superradiant response may be important.

Methodologically, the paper introduces a joint manipulated-lineshape polarimetry analysis for spin-1 AFP and hole-burned spectra and demonstrates it on an irradiated deuterated butanol AFP sequence. By combining shared global lineshape parameters with local rates-response and spin-temperature-consistency constraints, the method yields physically consistent component areas \(I^+\) and \(I^-\) even when AFP drives the system into non-Boltzmann half-flip or interrupted-sweep states. This enables extraction of both vector polarization \(P\) and tensor polarization \(Q\) from spectra for which standard equilibrium intensity-ratio methods are not valid.

Future work will focus on systematic mapping of \(\epsilon(A)\) at fixed and calibrated \(A\), direct proton and deuteron \(B_1\) calibrations, measurements of circuit \(Q\) and induced-field thresholds for AFP, and incorporation of the extracted \(P(t)\) and \(Q(t)\) into real-time target monitoring and feedback. Combinations of AFP with selective RF manipulation can also be studied for future experimental applications.

\section*{Acknowledgments}
This work was supported by DOE contract DE-FG02-96ER40950.

\appendix
\section{Uncertainty propagation}
\label{subsec:errors}
The statistical uncertainty assigned to each digitized point of an NMR spectrum is obtained from the noise level in the signal tails, where no absorption is present. Let $y_i$ denote the background-subtracted NMR signal amplitude at frequency point $R_i$. Points with $|R|>4$ are selected and a linear baseline $b(R)$ is fitted to remove small offsets or slopes. The noise level is then defined as the RMSD of the baseline-subtracted tail residuals,
\begin{equation}
  \sigma =
  \sqrt{\frac{1}{N_{\rm tail}}
  \sum_{|R_i|>4} \left(y_i - b(R_i)\right)^2},
  \label{eq:sigma_rmsd}
\end{equation}
where $N_{\rm tail}$ is the number of data points with $|R_i| > 4$.
This uniform $\sigma$ defines the statistical weight of each data point in the joint nonlinear least-squares fit of Sec.~\ref{sec:new_method}. For the irradiated deuterated butanol sequence analyzed in Sec.~\ref{sec:fit_results}, the calibration constant $C$ is determined from the final ten pre-AFP thermal-equilibrium spectra. Each spectrum $k$ provides an independent estimate $C_k$. The weighted mean calibration constant is
\begin{equation}
  C = \frac{\sum_k C_k/\sigma_k^2}{\sum_k 1/\sigma_k^2},
  \qquad
  \sigma_C =
  \left(\sum_k \frac{1}{\sigma_k^2}\right)^{-1/2}.
  \label{eq:C_weighted}
\end{equation}
The resulting $C \pm \sigma_C$ is applied to all spectra in the AFP sequence. For each AFP spectrum, the joint fit returns the best-fit parameter vector $\hat{\bm{\theta}}$ and covariance matrix $V_{\theta}$. The integrated component areas $I^+$ and $I^-$ are smooth functions of $\bm{\theta}$, so their covariance matrix is obtained from the Jacobian
\begin{equation}
V_I = J_I\,V_{\theta}\,J_I^{\mathsf T},
\qquad
J_I \equiv \frac{\partial (I^+,I^-)}{\partial \bm{\theta}}.
\label{eq:VI}
\end{equation}
Writing
\begin{equation}
V_I=
\begin{pmatrix}
\mathrm{Var}(I^+) & \mathrm{Cov}(I^+,I^-) \\
\mathrm{Cov}(I^+,I^-) & \mathrm{Var}(I^-)
\end{pmatrix},
\label{eq:VI_matrix}
\end{equation}
and neglecting correlations between $C$ and the AFP-fit parameters because $C$ is obtained from separate calibration spectra, the propagated variances of $P$ and $Q$ are
\begin{align}
\mathrm{Var}(P)
  &= (I^+ + I^-)^2\,\mathrm{Var}(C) \notag\\
  &\quad + C^2\!\left[\mathrm{Var}(I^+) + \mathrm{Var}(I^-) \right. \notag\\
  &\qquad\left. + 2\,\mathrm{Cov}(I^+,I^-)\right], \label{eq:varP} \\
\mathrm{Var}(Q)
  &= (I^+ - I^-)^2\,\mathrm{Var}(C) \notag\\
  &\quad + C^2\!\left[\mathrm{Var}(I^+) + \mathrm{Var}(I^-) \right. \notag\\
  &\qquad\left. - 2\,\mathrm{Cov}(I^+,I^-)\right]. \label{eq:varQ}
\end{align}

The calibration term dominates at large signals, while the relative uncertainty in $Q$ or $P$ increases as they approach zero. The covariance term may be retained explicitly to extract both areas from a single constrained fit. However, to remove the off-diagonal elements of $V_I$, $I^+$ and $I^-$ are obtained from independent fits to the spectrum~\cite{Keller2026}, each targeting one transition component. Since the fits share no parameters, their covariance matrices are decoupled by construction and $\operatorname{Cov}(I^+, I^-) = 0$ exactly. The quality of these fits is therefore limited by the constraints from the burn and AFP conditions, making accurate prior knowledge of the manipulations essential to minimize bias. Fitting each intensity separately can reduce the statistical error in particular circumstances but may introduce bias in others, so this method is used on a case-by-case basis.  Tests on simulated signals vs the fitted results are used to minimize systematic error in the method \cite{Keller2020,Keller2026}.

\bibliographystyle{elsarticle-num-names}
\biboptions{numbers,sort&compress}
\bibliography{references}

\end{document}